\documentclass[a4paper,preprint,superscriptaddress,preprintnumbers,nofootinbib]{revtex4}

\usepackage[dvipdf]{graphicx}
\usepackage{amsmath}
\usepackage{bm}
\usepackage{mathrsfs}

\newcommand{\Slash}[1]{{\ooalign{\hfil/\hfil\crcr$#1$}}}
\newcommand{\dcsb}{D$\chi$SB}
\newcommand{\ds}{\displaystyle}
\newcommand{\pa}{\partial}

\begin{document}
\title{Solving the QCD non-perturbative flow equation\\ as a partial differential 
equation and its application\\ to the dynamical chiral symmetry breaking}

\author{Ken-Ichi \surname{Aoki}}
\email{aoki@hep.s.kanazawa-u.ac.jp}
\affiliation{Institute for Theoretical Physics, Kanazawa University, Kanazawa 920-1192, Japan}

\author{Daisuke \surname{Sato}}
\email{satodai@hep.s.kanazawa-u.ac.jp}
\affiliation{Institute for Theoretical Physics, Kanazawa University, Kanazawa 920-1192, Japan}

\preprint{KANAZAWA-12-10}


\begin{abstract}
Non-perturbative renormalization group approach to the dynamical chiral symmetry breaking is an effective method which can accommodate beyond the ladder (mean filed) approximation. 
The usual method relying on the field operator expansion suffers explosive behaviors of the 4-fermi coupling constant, 
which prevent us from evaluating the physical quantities in the broken phase.
In order to overcome this difficulty, we solve the flow equation directly as a partial differential equation and calculate the dynamical mass and the chiral condensates. 
Also we formulate a beyond the ladder equation and it gives almost gauge independent results for the chiral condensates.
\end{abstract}

\maketitle

\section{Introduction}

The Wilsonian renormalization group approaches to the continuum quantum field theory have been formulated \cite{Wegner:1972ih,Polchinski:1983gv,Wetterich:1992yh} and developed in the past decades (see reviews \cite{Aoki:2000wm,Berges:2000ew,Pawlowski:2005xe,Gies:2006wv}).
In these approaches, the renormalization group (RG) flow equation is defined by 
a functional differential equation, 
the solution of which gives the partition function defined by the functional integral.
Using this framework, we are able to obtain new approximation methods extracting the non-perturbative information of the partition function.
Therefore we call this type of the RG ``non-perturbative renormalization group'' (NPRG).

In order to solve the flow equation approximately, we usually expand the equation in terms of the field operators and their derivatives.
The derivative expansion has been applied to evaluation of the universal 
quantities such as critical exponents and anomalous dimensions.
For example, in the three-dimensional scalar field theories, 
the expansion with respect to the field operator without the derivatives converges very well \cite{Tetradis:1993ts,Aoki:1996fn,Aoki:1998um,Canet:2002gs}.
Although it is difficult to confirm the convergence with respect to the order of
the derivatives, the result of the expansion up to the 4-th derivatives agrees well with the Monte Carlo simulations \cite{Canet:2003qd}.

The NPRG has been also applied to the analysis of the dynamical chiral symmetry breaking (\dcsb) in the strong coupling gauge theories and its effective theory.
As first noticed by Nambu and Jona-Lasinio \cite{Nambu:1961tp}, the scalar 4-fermi operators become the source of the \dcsb.
From the viewpoint of the NPRG, when we lower the renormalization scale of the effective action, 
the strong gauge interactions induce the effective 4-fermi operators, which bring about the {\dcsb} at the low energy scale \cite{Aoki:1996fh}.
Unless there are explicit symmetry breaking terms such as mass terms, the running 4-fermi coupling constant diverges at a low energy scale.
This explosive behavior is nothing but a signal of the \dcsb~\cite{Aoki:2000wm,Braun:2011pp}.

On the other hand, the (improved) ladder Schwinger-Dyson (SD) equation has been frequently used 
for the analysis of the \dcsb\ in the strong coupling gauge theories.
This ladder approximation has the strong dependence on the gauge-fixing parameter.
Moreover it is difficult to improve the ladder approximation systematically.
However the framework of NPRG allows us to take account of the effects of the non-ladder diagrams in a systematic fashion \cite{Aoki:1999dv,Aoki:2000dh}.

Since the running 4-fermi coupling constants diverge at a critical scale 
as mentioned above, 
the RG flow  can not go beyond the critical scale towards the infrared limit.
In the simple framework of NPRG which maintains the chirally symmetric structure of the effective action,
we can not evaluate the physical quantities in the broken phase such as the dynamical mass 
of the fermion and the condensates of the fermion bilinear composite operator.

To overcome this problem, we introduce the bare mass term, which also works as a source term of the chiral condensates \cite{Aoki:2009zza}.
We may expect that the bare mass prevents the divergent behavior of the 4-fermi coupling constants 
and might allow us to effectively evaluate the order parameter of the \dcsb\ at the infrared limit scale.
The field operator expansion of the NPRG equation, however, does not converge well at least
in the region of the bare mass as small as the current masses of up and down quarks.
Consequently we will take another way of directly solving the NPRG flow equation as a partial differential equation (PDE) without relying on any field operator expansion.

By the way, another method like the Hubbard--Stratonovich transformation has been 
used in many works to avoid the divergent behavior \cite{Aoki:1999dw,Gies:2001nw,Gies:2002hq}.
In this method, the composite operators of fermions are partially transformed to 
scalar fields, 
one of which obtains the nonzero expectation value as a chiral order parameter.
Introduction of those scalar fields has a merit that the meson physics can be argued 
simultaneously.
However it is more complicated to evaluate the convergence of the physical quantities 
with respect to the operator expansion including the scalar fields.

This article is organized as follows. In Sec.~2
we briefly  review the flow equation for the effective average action.
We introduce the basic truncation which projects the complete operator space 
onto the subspace relevant to the {\dcsb} so as to solve the flow equation 
approximately.
In Sec.~3 we examine the truncation method in detail and obtain two types of truncations:
one corresponds to the ladder approximation, and the other contains the main parts of the non-ladder corrections.
In Sec.~4 we explain how to evaluate the chiral order parameters.
In Sec.~5, we solve the flow equation by using the field operator expansion, and examine the convergence of chiral order parameters. 
In Sec.~6, we directly solve the partial differential equation and show the results of physical quantities.
In Sec.~7, we summarize our methods and results, and discuss further issues
along the line of thought of this article.

\section{NPRG flow equation and its application}
\label{sec-NPRG_flow_appli}
\subsection{Formulation}
In order to evaluate non-perturbative effects of the quantum field theory,
we introduce the so-called ``effective average action'' \cite{Wetterich:1992yh} that interpolates between the bare action and the full quantum effective action.
For this purpose, we define the generating function with the infrared cutoff 
supplied by the cutoff term $\Delta S_\Lambda[\Omega]$ as follows:
\begin{align}
 e^{W_\Lambda[J]}&\equiv Z_\Lambda[J]:=
 \int\! \mathcal{D}\Omega  e^{-S_{\rm bare}[\Omega]-\Delta S_\Lambda[\Omega]
 +\int\! J\cdot \Omega},
\end{align}
where $\Omega$ represents various fields generically and the functional integral is regularized by the ultraviolet (UV) momentum cutoff $\Lambda_0$.
The cutoff term $\Delta S_\Lambda$ is defined as the following mass term 
depending on the momentum: 
\begin{align}
 \Delta S_\Lambda [\Omega]&=\int_p \frac{1}{2}\Omega^T(-p)\cdot R_\Lambda(p)\cdot \Omega(p).
\end{align}
The regulator function $R_\Lambda(p)$ suppresses the quantum fluctuations with 
 the momentum lower than the infrared cutoff $\Lambda$.
Therefore the regulator function satisfies 
\begin{align}
 \lim_{q^2/\Lambda^2\rightarrow 0} R_\Lambda(p)>0\label{eq-cond_cut1} ,
\end{align}
and it implements the ``coarse graining'' in the Wilsonian method.

The effective average action is defined by a slightly modified Legendre transformation:
\begin{align}
 \Gamma_\Lambda[\Phi]&\equiv
 \sup_J \left(
  \int\Phi\cdot J -W_\Lambda[J]
 \right)
 -\Delta S_\Lambda[\Phi].
\end{align}
It satisfies the following boundary conditions,
\begin{align}
\begin{cases}
 \Gamma_\Lambda \rightarrow S_{\rm bare}, & \Lambda\rightarrow\Lambda_0\rightarrow\infty\label{eq-boundary_cond.} \\
 \Gamma_\Lambda \rightarrow \Gamma, & \Lambda \rightarrow 0,
\end{cases},
\end{align}
provided that the regulator function $R_\Lambda(p)$ has the following properties,
\begin{align}
\begin{cases}
 R_\Lambda(p)\rightarrow \infty,& \Lambda\rightarrow \Lambda_0 \rightarrow \infty
 \label{eq-cond_cut2}\\
 R_\Lambda(p)= 0, & \Lambda^2/q^2\rightarrow 0
\end{cases}.
\end{align}


The cutoff dependence of the effective average action is exactly given by the NPRG flow equation,
\begin{align}
 \partial_t \Gamma_\Lambda[\Phi]&=
 \frac{1}{2}{\rm STr}\left[\left(\Gamma^{(2)}_\Lambda+R_\Lambda\right)^{-1}\cdot \partial_t R_\Lambda\right],
 \label{eq-non_per_flow_eq}
\end{align}
where we define the dimensionless scale parameter $t=\log \Lambda_0/\Lambda$.
The right hand side should be called $\beta$ functional, and consists of the second order 
functional derivative of the effective average action,
\begin{align}
 \left(\Gamma_\Lambda^{(2)}\right)_{ij}(p,q)=\frac{\delta^2 \Gamma_\Lambda[\Phi]}{\delta\Phi_i(-p) \delta\Phi_j(q)}.
\end{align}
It is considered as a matrix with respect to the momentum and the species of fields.
This flow equation is firstly derived by Wetterich \cite{Wetterich:1992yh}.
Because of the boundary condition (\ref{eq-boundary_cond.}),
the flow equation interpolates between the bare action $S_{\rm bare}$ and the full quantum effective action $\Gamma$.

\subsection{Application to QCD}

Hereafter we consider the $N_{\rm f}$-flavor massless QCD with $N_{\rm c}$-color.
The bare action in Euclidean space is
\begin{align}
 S_{\rm bare}&=
 \int_x\!\left\{ \frac{1}{4}F^a_{\mu\nu}F^{\mu\nu}_a
 +\bar{\psi}\left(\Slash{\partial}+i\bar{g}_s\Slash{A}^aT^a\right)\psi
 \right\},
\end{align}
where $T^a$ is the generator of the fundamental representation of SU($N_{\rm c}$).
This action has the chiral symmetry $SU(N_{\rm f})_{\rm L}\times SU(N_{\rm f})_{\rm R}$, which is to be broken down to $SU(N_{\rm f})_{\rm V}$ dynamically by the strong gauge interactions.

To evaluate the dynamical chiral symmetry breaking (\dcsb) by using the non-perturbative renormalization group (NPRG) flow equation, we truncate field operators which are not 
essential to the {\dcsb} from the complete operator space of the effective average action.
Here we define the following truncated effective average action,
\begin{align}
 \Gamma_\Lambda[\Phi]&=
 \int_x\!\left\{ \frac{Z_F}{4}F^a_{\mu\nu}F^{\mu\nu}_a
 + \frac{1}{2\xi}\left(\partial_\mu A_\mu\right)^2
 +\bar{\psi}\left(Z_\psi\Slash{\partial}+i\bar{g}_s\Slash{A}\right)\psi
 -V(\psi,\bar{\psi};\Lambda)\right\},\label{eq-truncated_action}
\end{align}
where we use the covariant gauge with the gauge-fixing parameter $\xi$, and do not 
represent the ghost sector for simplicity.
This truncated subspace of the complete effective action is spanned by the operators of the bare QCD action and the fermion self-interaction operator $V(\psi,\bar{\psi};\Lambda)$, which we call the fermion potential.
The discarded operators including higher derivatives or gluon fields may affect 
quantitative evaluation, but are not the essential sector to drive the \dcsb.
Actually we will confirm that the operators of the bare QCD and the fermion potential bring about {\dcsb} using the NPRG flow equation \cite{Aoki:1999dw}.

Here it should be noted that the flow equation induces operators breaking the gauge symmetry, such as the mass operator of the gauge boson, because the cutoff function explicitly breaks the gauge symmetry.
However we do not discuss this issue because we truncate out such operators.
Eventually the flow equation of the gauge coupling constant agrees with that 
of the one-loop perturbation theory.
We will concentrate on evaluating the flow equation for the fermion potential.

\subsection{Techniques to derive the NPRG flow equation}
In the rest of this section, we will explain some techniques to explicitly write down the NPRG flow equation (\ref{eq-non_per_flow_eq}) for the fermion potential.
We denote the degrees of freedom of the fields as a vector $\Phi^t=(A_\mu^a, 
\psi^t,\bar{\psi})$.
Then the dressed inverse propagator including the cutoff function is
given by 
\begin{align}
\begin{split}
 \Gamma^{(2)}_\Lambda&[\Phi](p,q)+R_\Lambda(p,q)\label{eq-inverse_propagator}\\[5pt]
 &= 
\begin{pmatrix}
 \ds \frac{\overrightarrow{\delta}}{\delta A^a_\mu(-p)} \\[8pt]
 \ds \frac{\overrightarrow{\delta}}{\delta \psi(-p)} \\[8pt]
 \ds \frac{\overrightarrow{\delta}}{\delta \bar{\psi}^t(-p)}
\end{pmatrix}
 \Big(\Gamma_\Lambda[\Phi]+\Delta S_\Lambda[\Phi]\Big)
\begin{pmatrix}
 \ds \frac{\overleftarrow{\delta}}{\delta A^b_\nu(q)}, & 
 \ds \frac{\overleftarrow{\delta}}{\delta \psi^t(q)}, & 
 \ds \frac{\overleftarrow{\delta}}{\delta \bar{\psi}(q)}
\end{pmatrix}.
\end{split}
\end{align}
The regulator function is defined by
\begin{align}
 R_\Lambda(p,q)&=
 \begin{pmatrix}
  Z_A\ r(p)\  (D_0^{-1})^{ab}_{\mu\nu}(p)& 0 & 0 \\[2pt]
  0 & 0 & Z_\psi\ r_\psi(p)\  i\Slash{p}^t\\[2pt]
  0 & Z_\psi\ r_\psi(p)\ i\Slash{p} & 0
 \end{pmatrix}
 \times \delta(p-q),
\end{align}
where $(D_0^{-1})^{ab}_{\mu\nu}(p)\equiv p^2\delta^{ab}(\delta_{\mu\nu}-\frac{q_\mu q_\nu}{q^2}(1-\frac{1}{\xi}))$ and $\delta(p-q)\equiv(2\pi)^4\delta^{(4)}(p-q)$.
Functions, $r(p)$ and $r_\psi(p)$ are defined to satisfy the properties (\ref{eq-cond_cut1}) and (\ref{eq-cond_cut2}). 

Next we explain how to calculate the ``super trace'' in the NPRG flow equation (\ref{eq-non_per_flow_eq}).
We transform the flow equation as follows \cite{Berges:2000ew},
\begin{align}
 \partial_t \Gamma_\Lambda[\Phi]
 &=\tilde{\partial_t} \  \frac{1}{2} {\rm STr} \, 
   \log \left[ \Gamma^{(2)}_{\Lambda}+R_\Lambda \right].
 \label{eq-Str_NPRG_flow}
\end{align}
Here the symbol $\tilde{\partial_t}$ is defined by
\begin{align}
 \tilde{\partial}_t&= \int_p \bigg[\
 \frac{\partial_t(Z_A r(p))}{Z_A}\frac{\delta}{\delta r(p)}
 + \frac{\partial_t(Z_\psi r_\psi(p))}{Z_\psi}\frac{\delta}{\delta r_\psi(p)} \
 \bigg],
\end{align}
where $\int_p\equiv\int \frac{d^4p}{(2\pi)^4}$.
Then we split the inverse propagator matrix (\ref{eq-inverse_propagator}) into submatrices as follows,
\begin{align}
 M\equiv \Gamma^{(2)}_k+R_k&= 
\begin{pmatrix}
 M_{\rm BB}& M_{\rm BF}\label{eq_M}\\
 M_{\rm FB}& M_{\rm FF} 
\end{pmatrix},
\end{align}
where $M_{\rm BB}$ ($M_{\rm FF}$) corresponds to the second derivative with respect to bosonic (fermionic) fields, while $M_{\rm BF}$ and $M_{\rm FB}$ correspond to bosonic and fermionic field derivatives.
Physically the submatirx $M_{\rm BB}$ corresponds to the inverse free propagator of the gauge field, the submatrices $M_{\rm BF}$ and $M_{\rm FB}$ are the gauge interactions, and the submatrix $M_{\rm FF}$ contains the inverse propagator of the fermion and the fermion self-interactions.
Using this notation, we can rewrite the ``super-trace log'' in the flow equation (\ref{eq-Str_NPRG_flow}) into the following formula \cite{Clark:1992jr}:
\begin{align}
  {\rm STr} \log M &= -{\rm Tr} \log M_{\rm FF} 
  +{\rm Tr} \log \left[ M_{\rm BB}-M_{\rm BF}{M_{\rm FF}^{-1}M_{\rm FB}}\right].\label{eq-STr_log} 
\end{align}
Note that another expression, 
\begin{align}
{\rm STr} \log M={\rm Tr} \log M_{\rm BB} 
  +{\rm Tr} \log \left[ M_{\rm FB}-M_{\rm FB}{M_{\rm BB}^{-1}M_{\rm BF}}\right],
\end{align}
has been often used in the NPRG analyses.
Here Eq. (\ref{eq-STr_log}) is more appropriate for our purpose of improving 
the gauge parameter dependence.

To derive the flow equation for the fermion potential, 
the matrix $M$ in Eq. (\ref{eq_M}) should be evaluated by replacing the fields with their zero-momentum components. Replacing $\Phi(p)\rightarrow \Phi\delta(p)$, we have
\begin{align}
M&(p,q)=\Gamma^{(2)}_\Lambda[\Phi](p,q)+R_\Lambda(p,q)\Big|_{\Phi(p)=\Phi\delta(p)}
\notag \\[5pt]
\begin{split}
&=\begin{pmatrix}
 Z_F(1+r) (D_0^{-1})^{ab}_{\mu\nu}& i\bar{g}_s \bar{\psi}\gamma_\mu T^a & -i\bar{g}_s(\gamma_\mu T^a\psi)^T\label{eq-reduced_inv_pro} \\[2pt]
 -i\bar{g}_s (\bar{\psi}\gamma_\nu T^b)^t & -\dfrac{\pa^2 V}{\pa\psi\pa\psi} & iZ_\psi(1+r_\psi) \Slash{p}^t-\dfrac{\pa^2V}{\pa\psi\pa\bar{\psi}}\\[2pt]
 i\bar{g}_s \gamma_\nu T^b\psi & iZ_\psi(1+r_\psi)\Slash{p}-\dfrac{\pa^2V}{\pa\bar{\psi}\pa\psi}  & -\dfrac{\pa^2 V}{\pa\bar{\psi}\pa\bar{\psi}}
\end{pmatrix}\\
&\qquad\qquad\qquad\qquad\qquad\qquad\qquad\qquad\qquad\qquad\qquad\qquad\qquad\qquad
 \times \delta(p-q).
\end{split}
\end{align}

\section{Flow equation for the fermion potential}

\subsection{Scalar 4-fermi operator}
The gauge interactions induce all possible fermion operators respecting the symmetry of QCD, 
which are enhanced  by themselves when we lower the cutoff scale.
Even in the truncated subspace of the effective action (\ref{eq-truncated_action}),
we can not treat exactly all possible operators and interactions.
Therefore, as an approximation of the flow equation for the fermion potential,
we project its full operator space onto a specific subspace and restrict 
the interactions so that we evaluate the {\dcsb} most effectively.

As in the QED with one flavor \cite{Aoki:1999dv}, 
the central operator for {\dcsb} is undoubtedly the following scalar 4-fermi operator,
\begin{align}
 \rho&=\frac{1}{2} \sum_{I=0}^{N_{\rm f}^2-1}
       \left[
       \left( \bar{\psi}\lambda^I\psi\right)^2 +\left( \bar{\psi}\lambda^Ii\gamma_5\psi\right)^2 
       \right],
\end{align}
where $\lambda^I\ (I=1, \cdots, N_{\rm f}^2-1) $ are the generators of the fundamental representation of SU($N_{\rm f}$), 
and $\lambda^0=\frac{1}{\sqrt{2N_{\rm f}}}\, {\bf 1}_{\rm flavor}$ is defined so that 
they satisfy the proper normalization, ${\rm tr}[\lambda^I\lambda^J]=\frac{\delta^{IJ}}{2}$.
The 4-fermi operator $\rho$, often adopted in Nambu--Jona-Lasinio (NJL)-type model with $N_{\rm f}=3$, is  invariant under the chiral transformation, $SU(N_{\rm f})_{\rm L}\times SU(N_{\rm f})_{\rm R}$.
It is the only chiral invariant 4-fermi operator which gives corrections to 
the mass operator, 
and it becomes the relevant operator in the region of the strong gauge coupling 
constant.
Therefore, for a first-step approximation,
we project the operator space of the fermion potential onto 
the subspace spanned by polynomials in the scalar operator $\rho$.

To project the flow equation onto the subspace defined above,
we determine the coefficient from  all possible operators 
included in the full fermion potential. 
This is equivalent to count the coefficients of powers of 
$(\bar{\psi}\psi)^2$, even though this operator itself is not chirally invariant.
It is due to the fact that $(\bar{\psi}\psi)^2$ operator does not appear in chiral invariant operators other than powers of $\rho$.
Eventually, we may work with a potential function in the simplest scalar operator 
$\sigma=\bar{\psi}\psi$:
\begin{align}
 V(\psi,\bar{\psi})& \rightarrow V(\sigma).
\end{align}
Here we note that the original chiral symmetry is not maintained in this subspace,
but the discrete chiral symmetry still remains:  the Lagrangian (the fermion potential)
is invariant under the following discrete transformation,
\begin{align}
\psi\rightarrow\gamma_5 \psi,\ \bar{\psi}\rightarrow -\bar{\psi}\gamma_5,  \ 
\sigma \rightarrow -\sigma.\label{eq-disrete_chi_sym}
\end{align}
The discrete chiral symmetry forbids the operators of the odd powers of $\sigma$, 
such as a mass term.

Next we pick up the interactions that are expected to be most important for the {\dcsb} 
and for improvement of the gauge-fixing parameter ($\xi$) dependence.
In Eq. (\ref{eq-reduced_inv_pro}), we further select the large-$N_{\rm c}$ leading 
interactions in the fermion self-interactions, which leads to the following 
simplification:
\begin{align}
\begin{split}
 &\frac{\partial^2 V}{\partial\psi\partial\psi},\ 
 \frac{\partial^2 V}{\partial\bar{\psi}\partial\bar{\psi}} 
 \rightarrow
 0,\\
 &\frac{\partial V}{\partial\bar{\psi}\partial\psi},\
  -\frac{\partial V}{\partial\psi\partial\bar{\psi}}
 \rightarrow \partial_\sigma V.
\end{split}
\end{align}

Applying the above approximation and the usual field renormalization, 
$\psi\rightarrow \psi/Z_\psi^{1/2}$ and
 $A^a_\mu\rightarrow A^a_\mu/Z_A^{1/2}$, to Eq. (\ref{eq-Str_NPRG_flow}) and 
(\ref{eq-STr_log}),
we obtain the flow equation for the fermion potential,
\begin{align}
 \partial_t V(\sigma;t)&=
 -\eta_\psi \sigma\partial_\sigma V
 +\int_p\!{\rm tr}\, \tilde{\partial}_t \log S_\psi^{-1}(p)
 -\frac{1}{2} \int_p\! \tilde{\rm tr}\, \tilde{\partial}_t \log \left[1+A(p)+B(p)\right],\label{eq-general_flow_eq_potential}
\end{align}
where $\eta_\psi$ is the anomalous dimension of the fermion field.
Functions $A(p)$ and $B(p)$ are matrices in the space of the color adjoint representation and the Euclidean Lorentz vector space as follows:
\begin{align}
 (A)^{ab}_{\mu\nu}(p)&=g_s^2\bar{\psi}T^a\gamma_\mu S_\psi(p)\gamma_\rho D_{\rho\nu}(p)T^b\psi,\label{Eq-Matrix_A}\\
 (B)^{ab}_{\mu\nu}(p)&=g_s^2\bar{\psi}T^bD_{\nu\rho}(p)\gamma_\rho S_\psi(-p)\gamma_\mu T^a\psi.\label{Eq-Matrix_B}
\end{align}
Here we should pay attention to the difference between the two types of traces:
${\rm tr}$ acts the space of the fermion's representation, the Dirac spinor and the color fundamental representation and the flavor fundamental representation,
and $\tilde{\rm tr}$ acts the space of the matrix $A$, $B$.
The propagators of the fermions and the gauge bosons including the cutoff function in Eq. (\ref{Eq-Matrix_A}) and (\ref{Eq-Matrix_B}) are defined by
\begin{align}
 S_\psi(p)&= \frac{-i(1+r_\psi) \Slash{p}-\partial_\sigma V}{P_\psi(p)+\partial_\sigma V^2}, \\
 (D)^{ab}_{\mu\nu}(p)&= \frac{\delta^{ab}}{P(p)}
    \left(\delta_{\mu\nu}-(1-\xi)\frac{p_\mu p_\nu }{p^2}\right),
\end{align}
where $P_\psi(p)\equiv (1+r_\psi)^2 p^2$, and $P(p)\equiv (1+r)p^2$.
According to the field renormalization, the anomalous dimension $\eta_\psi( \equiv \partial_t \log Z_\psi )$ appears in Eq. (\ref{eq-general_flow_eq_potential}), and the gauge coupling constant is renormalized as $g_s= Z_1 \bar{g}_s/Z_\psi Z_A^{1/2}$.

In the next two subsections, we will explain how to extract the scalar operators from the right hand side of Eq. (\ref{eq-general_flow_eq_potential}).

\subsection{Ladder approximation}
\label{Subsec-ladder}
\begin{figure}[htbp]
 \center
 \includegraphics[width=0.7\hsize]{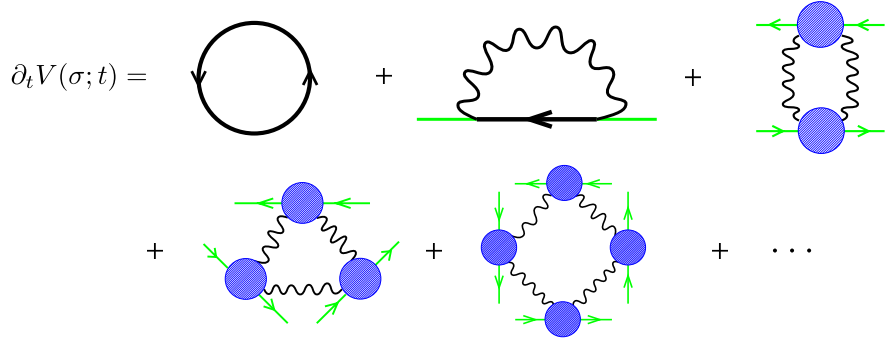}
 \caption{Infinite series of diagrams expressing the flow equation (\ref{eq-general_flow_eq_potential}) 
          using the corrected vertex.
          }
 \label{fig-ladder_form_diagrams_using_collected_vertex}
\end{figure}

\begin{figure}[htbp]
 \center
 \includegraphics[width=0.57\hsize]{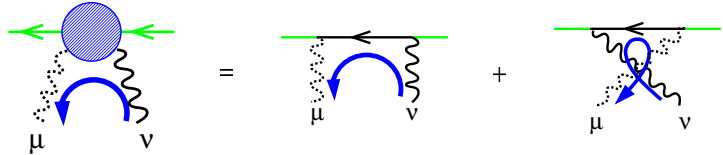}
 \caption{The corrected vertex. The two diagrams in the right hand side correspond to $A$ and $B$ in Eq.~(\ref{eq-general_flow_eq_potential}) respectively. The curved arrows denote the direction of the momentum flow.}
 \label{fig-A+B}
\end{figure}

In Fig. \ref{fig-ladder_form_diagrams_using_collected_vertex}, the flow equation (\ref{eq-general_flow_eq_potential}) is diagrammatically expressed by using the ``corrected vertex'' defined in Fig. \ref{fig-A+B}.
We regard $A$ and $B$ as the ladder element and the crossed element of the diagrams, respectively.

The diagrams consisting only of the ladder element $A$ are the ladder diagrams.
Moreover we find that the diagrams consisting  only of the crossed elements
 $B$ are also the ladder diagrams
if we rotate all the fermion lines of the diagrams to untangle the crossed gluon lines.
Therefore the ladder approximation with only the ladder diagrams should be defined by
\begin{align}
 -\frac{1}{2}\tilde{\rm tr}\log[1+A+B]
 &= \frac{1}{2}\sum_{n=1}^{\infty}\frac{(-1)^n}{n}\tilde{\rm tr} (A+B)^n \notag \\
 &\stackrel{{\rm ladder}}{\Longrightarrow} \frac{1}{2} \sum_{n=1}^{\infty}\frac{(-1)^n}{n}\tilde{\rm tr} (A^n + B^n)\notag \\
 &=  \sum_{n=1}^{\infty}\frac{(-1)^n}{n}\tilde{\rm tr} A^n.\label{eq-ladder_tr_log}
\end{align}

To project Eq. (\ref{eq-ladder_tr_log}) onto the subspace of the scalar operators $\sigma^n$,
we adopt the following approximation rule of picking up $\sigma$,
\begin{align}
 \bar{\psi}_1 O_1 \psi_2 \bar{\psi}_2 O_2 \psi_3 \cdots \bar{\psi}_n O_n \psi_1
 &\rightarrow
 (-1)^{n+1}F_n \prod_{i=1}^{n} \bar{\psi}_i {\bf 1}_{\rm spinor} \lambda^0 T^0 \psi_i,\label{eq-general_scalar_fierz}
\end{align}
where we use $T^0\equiv \frac{1}{\sqrt{2N_{\rm c}}}\, {\bf 1}_{\rm color}$.
The right hand side of the above formula is nothing but the scalar part of the general Fierz transformation obtained by using the completeness of the space of the spinor and the color and the flavor.
Thus the coefficient $F_n$ in the above formula is given by
\begin{align}
 F_n&={\rm tr} \left[{\bf 1}_{\rm spinor} \lambda^0 T^0 O_1 \cdot {\bf 1}_{\rm spinor} \lambda^0 T^0 O_2 \cdots {\bf 1}_{\rm spinor} \lambda^0T^0 O_n \right].
\end{align}

According to this rule, the summation in Eq.~(\ref{eq-ladder_tr_log}) is 
calculated as follows:
\begin{align}
 \sum_{n=1}^{\infty} \frac{(-1)^n}{n}\tilde{\rm tr}A^n&\rightarrow
 \sum_{n=1}^{\infty} {\rm tr}\left[\frac{(-1)^{n+1}}{n} 
   \left(-C_2\, g_{\rm s}^2\frac{3+\xi}{4P(p)}S_\psi(p)\, \frac{\sigma}{N_{\rm f}N_{\rm c}}
   \right)^n
 \right] \notag \\
 & ={\rm tr} \log\left(1-C_2\, g_{\rm s}^2\frac{3+\xi}{4P(p)}S_\psi(p)\, \frac{\sigma}{N_{\rm f}N_{\rm c}}
    \right),
\end{align}
where $C_2$ is the second Casimir invariant of the $SU(N_{\rm c})$ representation,
$C_2=\sum_{a=1}^{N_{\rm c}^2-1}T^aT^a$.

Finally we obtain the ladder flow equation, 
\begin{align}
 \partial_t V(\sigma;t)&=
 -\eta_\psi \sigma\partial_\sigma V
 +\int_p\!{\rm tr}\, \tilde{\partial}_t
  \log \left(
   S_\psi^{-1}(p) -C_2\, g_{\rm s}^2 \frac{3+\xi}{4P(p)} \sigma
  \right),\label{Eq-general-ladder_flow_eq}
\end{align}
where we rescaled $V$ and $\sigma$ by common factor $N_{\rm c}N_{\rm f}$,
$V\rightarrow N_{\rm c}N_{\rm f} \, V$ and $\sigma \rightarrow N_{\rm c}N_{\rm f} \, \sigma$.

As for the regulator function, we adopt the following sharp regulator function,
\begin{align}
r_\mathrm{sharp}(p)= Z_A \cdot r(p)  =Z_\psi \cdot r_\psi(p)=\frac{1}{\theta(p^2-1)}-1.\label{eq-sharp_cutoff_function}
\end{align}
Performing the momentum integration in Eq. (\ref{Eq-general-ladder_flow_eq}), we obtain the ladder flow equation as a partial differential equation (PDE),
\begin{align}
 \partial_t V(\sigma;t) &=
 -\eta_\psi \sigma\partial_\sigma V+
\frac{\Lambda^4}{4\pi^2}\ln\left[1+
   \Lambda^{-2}\left(
      \partial_\sigma V+(3+\xi)\frac{ C_2\, g^2_{\rm s}\sigma}{4\Lambda^2}
               \right)^2
   \right].\label{eq-ladder-flow-equation}
\end{align}
Apart from the anomalous dimension term,
this flow equation agrees with the local potential approximated Wegner-Houghton equation,
and it was proven that the flow equation gives the results equivalent to the improved ladder Schwinger-Dyson equation \cite{Aoki:1999dw}.
Actually this is the reason why we call this approximated flow equation ``the ladder''.

Using the momentum scale expansion \cite{Morris:1993qb,Morris:1995af}, the anomalous dimension of the fermion field $\eta_\psi=\partial_t \log Z_\psi$ is given by
\begin{align}
 \eta_\psi&=\frac{C_2 g^2_{\rm s}}{8\pi^2}
 \left(\xi\frac{\Lambda^2}{\Lambda^2+m^2(t)}+\frac{3-\xi}{4}\frac{\Lambda^2m^2(t)}{(\Lambda^2+m^2(t))^2}\right),
\end{align}
where we define the running mass,
\begin{align}
 m(t)= \partial_\sigma V(\sigma;t)|_{\sigma=0}\, .
\end{align} 

As will be seen in the numerical results in Sec. \ref{Subsec-solving_PDE}, the chiral order parameters given by the ladder flow equation (\ref{eq-ladder-flow-equation}) strongly depend on the gauge-fixing parameter $\xi$.

\subsection{Beyond the ladder approximation}

In order to improve the gauge dependence, we have to add the crossed element as well. 
We evaluate the flow equation (\ref{eq-general_flow_eq_potential}) using the full corrected 
vertex in Fig.~\ref{fig-A+B}, which is calculated as follows:
\begin{align}
\begin{split}
 (A+B)^{ab}_{\mu\nu}(p)&=
 -\frac{2g_s^2}{P_\psi+\partial_\sigma V^2}\bar{\psi}T^aT^b\left(
 i\frac{(1+r_\psi)p_\alpha}{P}(\epsilon_{\mu\nu\alpha\beta}\gamma_5\gamma_\beta)+MD_{\mu\nu}
 \right)\psi\label{eq-A+B_abelian_commutator} \\
&\quad  +\text{(term including }[T^a,T^b] \text{ )}.
\end{split}
\end{align}
Here we will ignore the commutator term for simplicity.
Note that this ignorance is consistent with our approximation that we do not include 
diagrams with gluon self couplings.
From the following discussion we expect that this ignorance does not induce the strong gauge dependence in the truncated subspace.

We associate the corrected vertex with the gauge independent set of diagrams for the S-matrix in case of the Abelian gauge theory.
In the non-Abelian gauge theory,
the diagram exchanging one gluon is also needed for the gauge independence of the S-matrix.
On the other hand, such corrections can not be added directly due to the one-loop nature of the NPRG $\beta$ function,
and therefore the gauge dependence which would be canceled in the S-matrix appears in the commutator term of Eq. (\ref{eq-A+B_abelian_commutator}).
In the NPRG, the correction by the diagrams exchanging one gluon is treated through
the effective operators such as $\partial_\mu F_{\mu\nu}\bar{\psi}\gamma_\nu\psi$.
However such operators are not included in the truncated subspace.
Thus we may omit the commutator term without suffering strong gauge dependence.

Then, according to the general Fierz transformation (\ref{eq-general_scalar_fierz}), we can pick up the scalar operators as follows:
\begin{align}
\begin{split}
\frac{(-1)^n}{2n}\tilde{\rm tr}(A+B)^n  
 &\rightarrow 
  2N_{\rm c}N_{\rm f} \frac{(-1)^{n+1}}{n}
  \left[
  \frac{2g_{\rm s}^2}{2P(P_\psi+M^2)}\frac{\sigma}{N_{\rm c}N_{\rm f}}
  \right]^n \\
  &\qquad \times\left[
   (\xi M)^n
  +\sum_{k=0}^{[\frac{n}{2}]}(-1)^k{{n}\choose{2k}}(2+4^k)
   M^{n-2k}{P_\psi}^{k}
  \right].
\end{split}
\end{align}
Finally, adopting the sharp regulator function, we obtain the flow equation beyond the ladder approximation as the following partial differential equation:
\begin{align}
\begin{split}
 \partial_t V(\sigma;t)&= 
 -\eta_\psi \sigma\partial_\sigma V+
 \frac{\Lambda^4}{4\pi^2}\log\left[1+\frac{B^2}{\Lambda^2}\right]
   +\frac{\Lambda^4}{8\pi^2}
     \log\left[
      \frac{\Lambda^2+B^2}{\Lambda^2+\partial_\sigma V^2}
      +\frac{3\Lambda^2G^2}{(\Lambda^2+\partial_\sigma V^2)^2}
     \right]\label{eq-nonladder_flow_eq} \\
   &\quad +\frac{\Lambda^4}{4\pi^2}
     \log\left[
      1+\xi\frac{\partial_\sigma V\, G}{\Lambda^2+\partial_\sigma V^2}
     \right],
\end{split}
\end{align}
where $B=\partial_\sigma V+C_2\frac{g_s^2\sigma}{2\Lambda^2}$ and $G=C_2 \frac{g_s^2\sigma}{2\Lambda^2}$.

\section{Chiral order parameters}\label{subsec-4.1}
Now we explain how to evaluate the two chiral order parameters,
the dynamical mass of quarks and the chiral condensates 
$\langle\bar{\psi}\psi\rangle$, which are generated by the {\dcsb}.

In the framework of the NPRG, calculating the non-zero chiral order 
parameter is nontrivial because the NPRG flow equation maintains 
the chiral invariant structure of the effective action, which forbids appearance of the dynamical mass operator.
\footnote{
In the theories including the scalar fields whose symmetries spontaneously 
break, we can evaluate the nonzero expectation values of the scalar fields as 
the order parameters  by searching the minimum point of its effective potential.
In the theory we consider,
this method cannot be used directly because the chiral order parameters are not
 such expectation values of the fields.
Therefore, the scalar fields  
corresponding to the $\bar{\psi}\psi$ are introduced by methods like the 
Hubbard-Stratonovich transformation in many works.
In this article, however, we adopt another method as will be seen in 
Sec.~\ref{Subsec-solving_PDE}.
}

On the other hand the {\dcsb} shows itself as a divergent behavior of 
the 4-fermi coupling constant, which is the source of the {\dcsb} in the NJL 
model.
We can define the $\beta$ function for each operator by expanding the flow equation 
in powers of $\sigma$.
The $\beta$ function for the 4-fermi coupling constant consists of itself and 
the gauge coupling constant, but does not include the higher dimensional 
operators due to the chiral invariance.
Solving the RG equation, we obtain the flow of the 4-fermi coupling 
constant as follows:
Lowering the cutoff scale $\Lambda(t)$,
the gauge interactions generate the 4-fermi operator,
which enhances itself, and consequently the 4-fermi coupling constant
diverges at a finite infrared scale $\Lambda_{\rm c}$.

Because of this divergence, the RG flow can not go beyond the critical scale 
$\Lambda_{\rm c}$ toward the infrared limit in the chiral invariant operator space. 
Therefore the divergence seems to imply that the chiral invariant RG flow cannot exist at the cutoff scale lower than $\Lambda_{\rm c}$,
where the true RG flow might be in the chiral variant operator space including the mass operator.
The relation between the divergence and the {\dcsb} has been discussed 
in Refs.~\cite{Aoki:2000wm,Aoki:1999dv,Gies:2005as}.

Here, in order to go beyond the critical scale $\Lambda_{\rm c}$
and to effectively evaluate the chiral order parameters,
we introduce the bare mass term, which explicitly breaks the chiral symmetry, 
in addition to the chiral invariant bare action: \cite{Aoki:2009zza}
\begin{align}
S_{\rm bare}=S_{\rm bare}({\rm invariant})-\int_x m_0\, \bar{\psi}\psi.
\end{align}
When the cutoff scale $\Lambda(t)$ lowers,
the running mass $m(m_0;t)$, being $m_0$ at the initial scale $t=0$, is rapidly enhanced around $\Lambda_{\rm c}$ by the 4-fermi interaction,
and consequently the enhanced mass suppresses the $\beta$ functions due to the decoupling effect.
Therefore the 4-fermi coupling constant is expected  to stay finite at the infrared scale.
Taking the zero bare mass limit after solving the flow equation, 
the infrared limit mass becomes a chiral order parameter, the so-called dynamical mass of quarks:
\begin{align}
 m_{\rm dyn}&\equiv \lim_{m_0\rightarrow +0} \lim_{t\rightarrow \infty}
 m(m_0;t).\label{eq-m_dyn}
\end{align}

The bare mass $m_0$ also works as an external source for the composite operator $\bar{\psi}\psi$, and its expectation value is evaluated as follows:
\begin{align}
 \langle\bar{\psi}\psi\rangle&\equiv
 \lim_{m_0\rightarrow +0} \lim_{t\rightarrow \infty}
 \frac{\partial \bar{G}_0(m_0;t)}{\partial m_0},\label{eq-chi_cond}
\end{align}
where $\bar{G}_0(m_0;t)$ denotes $V(\psi,\bar{\psi};t)|_{\psi=\bar{\psi}=0}$.
Here the infrared limit value $\bar{G}_0(m_0;\infty)$ corresponds to the Helmholtz free energy.

From the view point of the Helmholtz free energy,
we may describe the relation between the {\dcsb} and the divergent behavior of the 4-fermi coupling constant.
Due to the chiral invariance of the theory, $\bar{G}_0(m_0;t)$ is an even function of $m_0$.
Hence the derivative of $\bar{G}_0(m_0;t)$ at $m_0=0$, namely the chiral condensates, vanishes if it is an analytic function.
On the other hand, the derivative of $\bar{G}_0(m_0;t)$ cannot be continuous at $m_0=0$, if the chiral condensates has the non-vanishing values.
When the cutoff scale $\Lambda(t)$ decreases,
the initially analytic function $\bar{G}_0(m_0;t)$ changes into a non-analytic one  at the critical scale $\Lambda_{\rm c}$.
Therefore its second derivative diverges at the scale $\Lambda_{\rm c}$.
Actually, in the large-N approximated NJL model, the 4-fermi coupling constant is almost equal to the second derivative, and therefore its divergent behavior shows the emergence of the {\dcsb} at that scale $\Lambda_{\rm c}$.

\section{Field operator expansion and its convergence}\label{subsec-4.2}
Here we attempt to solve the flow equation using the field operator
expansion so that we evaluate the chiral order parameters introduced 
in the previous section.
As seen in the end of this section, however, the field operator expansion 
does not work well in this model.

In the subspace spanned by polynomials in $\sigma$,
the $\beta$ function for the 4-fermi coupling constant includes the 6-fermi coupling constant owing to the chiral symmetry breaking effect of the bare mass $m_0$.
In general, the $\beta$ function for the $2n$-fermi coupling constant includes the $2(n+1)$-fermi coupling constant.
Therefore we have to take into account of the infinite number of the $2n$-fermi operators, and encounter the infinite tower of the coupled RG equations.

Since such infinitely coupled equations can not be evaluated numerically,
we stop the expansion at some maximum power $N$:
\begin{align}
V(\sigma ;t) = 
\sum_{n=0}^{N}\frac{1}{n!}  \bar{G}_n(t)\sigma^n,\label{eq-expand_zero}
\end{align}
where we call $N$ the order of ``truncation''.
Then, expanding the flow equation in powers of $\sigma$,
we set up $(N+1)$-coupled RG equations where the $\beta$ function 
for the $2N$-fermi operator does not include the $2(N+1)$-fermi coupling constant.
The lager the truncation order $N$ is, the better the solution of the truncated
 coupled RG equations approximates the solution of the original flow equation.
Actually, this type of truncation approximation has worked well in many theories.

\begin{figure}[h]
\center
\includegraphics[width=0.6\hsize]{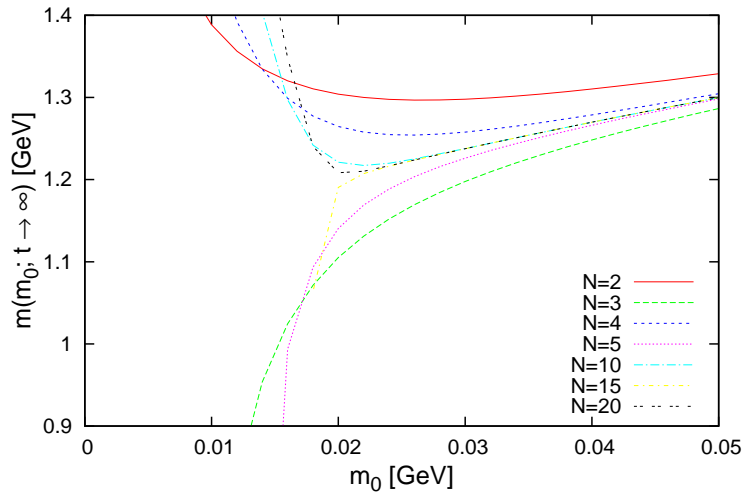}
\caption{Dependence of the infrared limit of the running mass $m(m_0;t\rightarrow \infty)$
on the bare mass $m_0$ and truncation order $N$.}
\label{fig-m_dyn_VS_m0}
\end{figure}

Let us see the truncation dependence of the dynamical mass 
$m_{\rm dyn}$ calculated for the ladder flow equation 
(\ref{eq-ladder-flow-equation}).
In  App. \ref{Sec-input_parameter}  we explain 
the input parameters and the running gauge coupling constant which obeys the one-loop perturbative $\beta$ function.
In Fig. \ref{fig-m_dyn_VS_m0}, we plot the running mass $m(m_0;t)$ 
at the infrared limit ( $t\rightarrow\infty$) for each bare mass,
and show its dependence on the truncation order from $N=2$ to $N=20$.
In the bare mass larger than about $0.02$ GeV,
the truncated solutions converge well with respect to truncation order $N$.
It also shows that the mass is dynamically generated by the {\dcsb} since 
the infrared running mass is much larger than the corresponding bare mass.
In the small bare mass region, however, the truncated solutions do not converge but diverge more badly for  larger truncation order.
Therefore we can not take the zero bare mass limit straightforwardly in this method with the field operator expansion.
It is also difficult to make a reliable extrapolation.
The non-ladder flow equation (\ref{eq-nonladder_flow_eq}) shows the similar behaviors as the ladder flow equation.

\begin{figure}[h]
 \center
 \includegraphics[width=0.6\hsize]{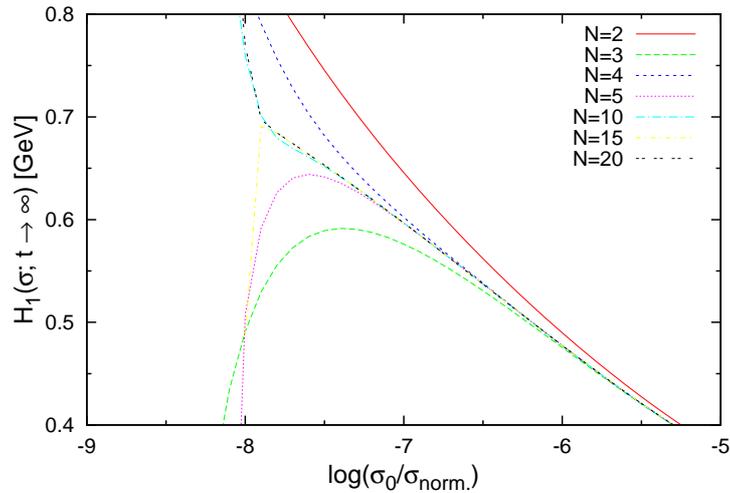}
 \caption{Dependence of $\bar{H}_1(\sigma_0;t\rightarrow \infty)$ on the expansion point $\sigma_0$ and truncation order $N$, where the normalization scale of $\sigma_0$ is  $\sigma_{\rm norm.}=1\ {\rm (GeV)^3}$.}
 \label{fig-sig0_H}
\end{figure}

Here we show the results using another method of avoiding the explosive behavior  of 4-fermi coupling constant.
In Eq. (\ref{eq-expand_zero}), the fermion potential is expanded around the 
vanishing value ($\sigma=0$).
We expand it around the nonvanishing value $\sigma_0$ ($\neq 0$),
\begin{align}
V(\sigma;t) = 
\sum_{n=0}^{N}\frac{1}{n!}  \bar{H}_n(k)
(\sigma-\sigma_0)^n.
\end{align}
In this expansion, we can go beyond the critical scale $\Lambda_{\rm c}$ without introducing the bare mass 
because the nonvanishing value $\sigma_0$ plays a similar role as the bare mass.
Note that it does not work as an external source for the chiral condensates.
The dynamical mass corresponding to Eq. (\ref{eq-m_dyn}) is given by the following limit,
\begin{align}
 m_{\rm dyn}& = \lim_{\sigma_0\rightarrow +0}\lim_{t\rightarrow \infty}
                \bar{H}_1(\sigma_0;t).\label{eq_m_dyn_sig0}
\end{align}
As seen in Fig. \ref{fig-sig0_H}, however, the expansion around the 
nonvanishing value does not converge in the small value of the expansion
 point $\sigma_0$.

\section{Solving the flow equation as PDE}
\label{Subsec-solving_PDE}

In Sec. \ref{subsec-4.1} and \ref{subsec-4.2}, so as to evaluate the chiral
order parameters, we have introduced two methods, introduction of
the bare mass and expansion around the non-zero point, both of which are expected
to avoid the divergent behavior of the 4-fermi coupling constant.
Note that the expansion around non-zero point means that the fermion potential 
$V(\sigma;t)$ is evaluated at the non-zero point, $\sigma\neq 0$.
In these methods, however, the field operator expansion does not converge
in the small value region of the bare mass or the expansion point, 
and we cannot obtain reliable chiral limit of the dynamical mass.

Obviously the poor convergence of the field operator expansion originates from the divergent behavior of the 4-fermi coupling constant,
which corresponds to the second derivative of the fermion potential 
$V(\sigma;t)$.
On the other hand, if the flow equation is solved as a partial differential equation (PDE), 
the fermion potential is expected to behave analytically at least except for the origin.
\footnote{
Strictly speaking, such a function without the total analyticity cannot be a global solution of the PDE.
Adopting the ``weak'' solution of the PDE, we can exactly solve the PDE 
and obtain physical quantities straightforwardly without any extrapolation.
This method of the weak solution will be reported in a separate article \cite{Aoki:2013}.
}
Therefore, we will directly solve the flow equation as a PDE without the field operator expansion.

In the practical calculation we solve the flow equation in terms of the mass
function, $M(\sigma;t)=\partial_\sigma V(\sigma;t)$, because the numerical 
solution of the PDE for the mass function is more stable than that for 
 the fermion potential.
Moreover, for the numerical stability around $\sigma=0$, $\sigma$ is 
transformed into the logarithmic variable, $x=\log \sigma/\sigma_{\rm norm}$
where the normalization scale $\sigma_{\rm norm}$ is set to be $1\ (\rm GeV)^3$.
We do not directly treat the origin of $\sigma$ and consider the limit $x\rightarrow -\infty$ of the mass function $M(x;t)$ to calculate the dynamical mass as seen in Eq. (\ref{eq_m_dyn_sig0}).

Now we adopt the simple formulation of the grid method where the derivatives with respect to $x$ are replaced with the finite differences.
The finite differences are defined by the 7-point formula where for example
the finite difference corresponding to the first derivative $\partial_x M$ 
consists of values on 6 points.
These points are nearest neighbors to the point where the derivative is 
evaluated.

For the numerical calculation, we will set a finite region for $x$, $x_{\rm L}\leq x\leq x_{\rm R}$.
In order to approximate the mass function at the end point $x_{\rm L}$ to be the dynamical mass, we need to choose a small enough $x_{\rm L}<<-1$.
Near the boundaries, where we can not take full 6 points, we use the 5-point formula, the 3-point formula, and eventually at the very end we take only the next point.
Usually these constraints enhance the numerical error near boundaries.
It will be found that actually these low order approximated definition of the
 finite difference near the boundaries does not induce the instability.
In the practical calculation we choose the boundaries such that $x_{\rm L}=-19$ and $x_{\rm R}=0$.

\begin{figure}[h]
 \center
 \includegraphics[width=0.7\hsize]{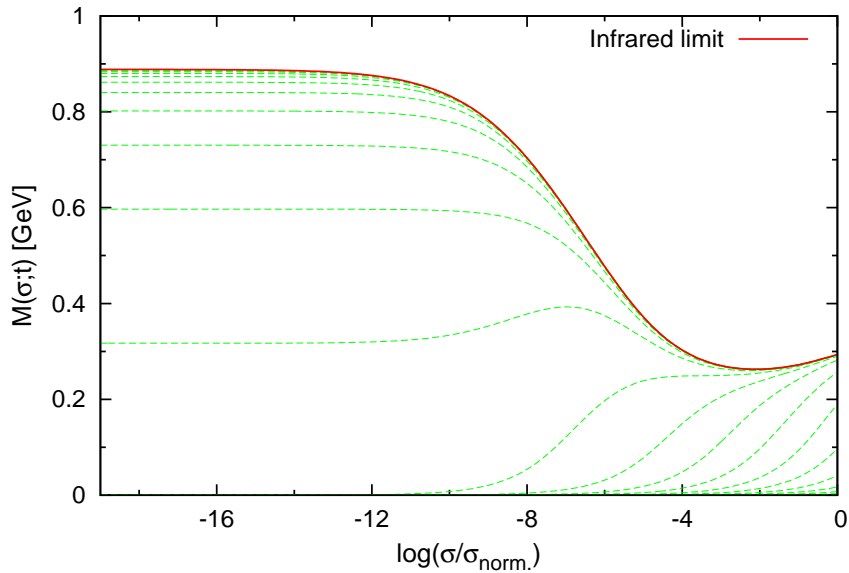}
 \caption{
 RG evolution of the mass function $M(\sigma;t)$. 
 The green dash lines denote the mass  function at dimensionless scales $t$ increasing by $\Delta t=0.3$ step,
 and the red line denotes the mass  function in the infrared limit, $t\rightarrow \infty$.
 The mass function is dramatically increased from the vanishing value below a critical scale $t_{\rm c}\simeq 5.3 $.}
 \label{fig-flow_mass_function}
\end{figure}

Through the procedure explained above, the coupled ordinary differential 
equations
for the discretized mass function on the grid are obtained,
and we numerically solve the equations with respect to the dimensionless scale
$t$ using the 4-th order Runge-Kutta method.
In Fig. \ref{fig-flow_mass_function} we present the RG evolution of the mass 
function given by the ladder flow equation (\ref{eq-ladder-flow-equation}).
Here the Landau gauge $\xi=0$ is adopted,  and the anomalous dimension is ignored,
which is consistent with the local potential approximation (LPA).
We can see that the mass function is dramatically increased from the vanishing 
value when lowering the cutoff scale $\Lambda(t)$.
Particularly the dynamical mass generation is observed below a critical scale,
rather rapidly in a short range of scale $t$.
The infrared limit is reliably evaluated, whose size reaches the $\Lambda_{\rm QCD}$ scale, and thus the {\dcsb} occurs. 
If the chiral symmetry is not spontaneously broken, the mass function goes to be zero in the limit $x\rightarrow -\infty$.

The mass function well converges to a certain value when $x$ goes towards the end point $x_{\rm L}$.
Therefore we can conclude that the approximated definition of the finite 
differences near the boundaries does not induce the instability of the mass function
at $x_{\rm L}$, and the value can be identified with the dynamical mass 
$m_{\rm dyn}$.
To evaluate the chiral condensates $\langle \bar{\psi}\psi\rangle$ by using Eq.~(\ref{eq-chi_cond}), we introduce the bare mass and calculate the free energy $\bar{G}_0$ through the evolution of $V$ at the end point $x_{\rm L}$.
As mentioned in Sec. \ref{Subsec-ladder}, the LPA ladder flow equation with the
Landau gauge has been proved to give the result equivalent to the improved ladder SD equation.
Actually the two chiral order parameter obtained now agree well with the ones
 obtained by the SD approach in Ref.~\cite{Aoki:1990eq},
which assures the total consistency of our method.

We present the numerical results of the two chiral order parameters 
obtained from the two approximated flow equations, the ladder one 
(\ref{eq-ladder-flow-equation}) and the non-ladder one 
(\ref{eq-nonladder_flow_eq}), with the various values of gauge fixing parameter $\xi$.
Table~\ref{tab-m_dyn} (Table~\ref{tab-chi_cond}) shows the numerical values of the 
dynamical mass (the chiral condensates) with or without the anomalous dimension.
Here the chiral condensates are the renormalized ones at $1\ \rm GeV$,
$\langle \bar{\psi}\psi\rangle_{\rm 1GeV}$ \cite{Aoki:1990eq}.
\begin{table}[h]
\center
 \caption{Results of the dynamical mass $m_{\rm dyn.}\ {\rm [GeV]}$ .
(L) ladder approximation; (NL) non-ladder approximation; (LA) ladder approximation with the anomalous dimension; 
(NLA) non-ladder approximation with the anomalous dimension. }
 \label{tab-m_dyn}
\begin{tabular}[t]{ccccc}
 \hline\hline
 $\xi$ & (L) & (NL) & (LA) & (NLA) \\
 \hline
 0 & 0.888 & 0.958 & 0.792 & 0.860 \\
 1 & 1.12  & 1.14  & 0.827 & 0.860 \\
 2 & 1.33  & 1.30  & 0.861 & 0.844 \\
 3 & 1.54  & 1.44  & 0.896 & 0.821 \\
 4 & --    & --    & 0.931 & 0.792 \\
 5 & --    & --    & 0.967 & 0.757 \\
 \hline\hline
\end{tabular}
\end{table}
\begin{table}[h]
 \center
 \caption{Results of the chiral condensates 
 $(\langle\bar{\psi}\psi\rangle_{\rm 1 GeV}/N_{\rm f})^{1/3}$ [GeV]. 
 The notation is the same as in  Table~\ref{tab-m_dyn}.}
 \label{tab-chi_cond}
\begin{tabular}[t]{ccccc}
 \hline\hline
 $\xi$ & (L) & (NL) & (LA) & (NLA) \\
 \hline
 0 & 0.215 & 0.219 & 0.209 & 0.213 \\
 1 & 0.267 & 0.264 & 0.221 & 0.219 \\
 2 & 0.330 & 0.312 & 0.236 & 0.222 \\
 3 & 0.408 & 0.364 & 0.255 & 0.224 \\
 4 & --    & --    & 0.278 & 0.225 \\
 5 & --    & --    & 0.304 & 0.225 \\
 \hline\hline
\end{tabular}
\end{table}

\begin{figure}[h]
 \center
 \includegraphics[width=0.6\hsize]{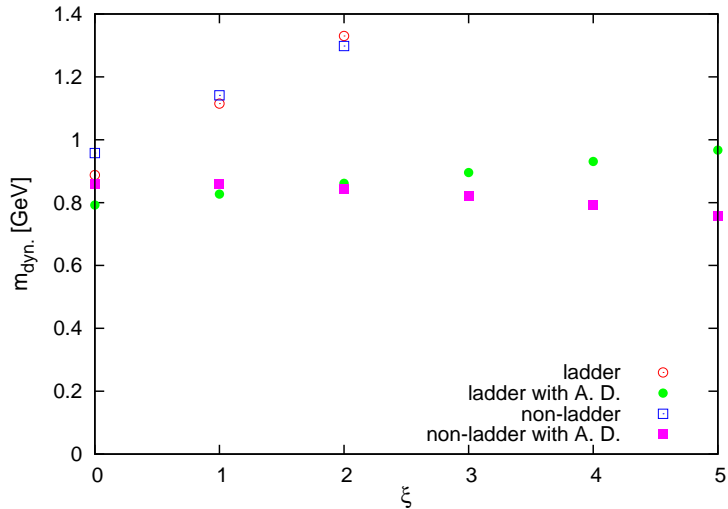}
 \caption{Gauge dependence of the dynamical mass $m_{\rm dyn}$. }
 \label{fig-m_dyn_xi}
\end{figure}

\begin{figure}[h]
 \center
 \includegraphics[width=0.6\hsize]{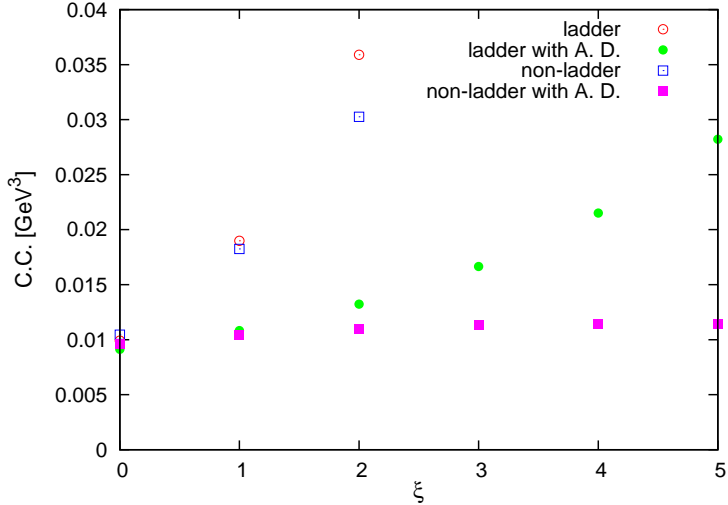}
 \caption{Gauge dependence of the chiral condensates (C. C.) 
 $\langle\bar{\psi}\psi\rangle_{\rm 1 GeV}/N_{\rm f}$.}
 \label{fig-chi_cond_xi}
\end{figure}

We show these results graphically in Fig. \ref{fig-m_dyn_xi} and 
\ref{fig-chi_cond_xi}.
Now we find that the gauge dependence of the results with the 
anomalous dimension (A. D.) are suppressed much better than ones without the A. D..
Besides, as for the results with the A. D., the gauge dependence 
of the chiral condensates obtained from the non-ladder flow equation 
is almost vanishing.
It looks perfect and this is nothing but what we expected to get by adding the crossed gluon diagrams to go beyond the ladder.
On the other hand, such great improvement by the non-ladder calculation is not seen for the dynamical mass.
This is due to the fact that the chiral condensates are the on-shell quantities while the dynamical mass is not.
Consequently there is no reason that the dynamical mass does not depend on the gauge fixing parameter.
We confirmed that our non-ladder extended calculation respects the gauge independence almost perfectly. 

Finally it should be noted that in the Landau gauge ($\xi=0$) the gauge dependent ladder result of the chiral condensates coincides with the almost gauge independent non-ladder extended one.
This feature of the Landau gauge proves a folklore that the ladder approximation is good particularly in the Landau gauge.

\section{Summary and discussion}
In this paper we have derived a new approximated flow equation beyond the ladder 
approximation so as to improve the dependence on the gauge fixing parameter.
We have developed various methods to evaluate the two chiral order parameters, the dynamical mass of quarks and the chiral condensates.
These methods are expected to avoid the explosive behavior of the 4-fermi coupling constant in the course of solving the flow equation. 
Within these methods, however, the field operator expansion, 
which is usually applied to solve the flow equations, shows poor convergence induced by the infrared singularity.
Then, we stopped the field operator expansion, and solved the flow equation as a partial differential equation by using the grid method.
We have obtained the chiral order parameters successfully without any instability or extrapolation.

As for the chiral condensates, we have seen that the gauge dependence of the non-ladder extended result almost disappeared.
Therefore our non-ladder extended approximation almost respect the gauge invariance.

A next step for further improvement of approximation would be to take into account of higher order operators including the first derivatives.
This improvement is implemented by replacing the coefficient of the kinetic term $Z_\psi(t)$ with the function of $\sigma$, $Z_\psi(\sigma;t)$.
Consequently we need to solve coupled partial differential equations in terms of the fermion potential $V(\sigma;t)$ and the kinetic function $Z_\psi(\sigma;t)$.

Also we have seen that particularly in the Landau gauge the gauge dependent ladder result of
the chiral condensates coincides with the almost gauge independent non-ladder extended one.
This agreement might be related to a statement that only in the Landau gauge the ladder approximation satisfies the Ward-Takahashi identity.
However, at finite temperature and chemical potential, this relation is broken.
Therefore we would encounter the new phase structures by applying this 
non-ladder extended approximation to the hot and dense QCD in the framework of the 
non-perturbative renormalization group.

\section*{Acknowledgements}
We would like to thank K. Miyashita and Y. Fujii for valuable  discussions.

\appendix 

\section{Running gauge coupling constant and input parameter}
\label{Sec-input_parameter}

In our truncated subspace, the $\beta$ function for the gauge coupling 
constant, $\alpha_{\rm s}\equiv g_{\rm s}^2/4\pi$, agrees with the result of the 1-loop perturbation:
\begin{align}
 \partial_t \alpha_{\rm s}&= \frac{\beta_0}{2\pi}\alpha_{\rm s}^2,
\end{align}
where $\beta_0=\frac{11}{3}N_{\rm c}-\frac{2}{3}N_{\rm f}$.
The solution of the RG equation is 
\begin{align}
 \alpha_{\rm s}&=\frac{2\pi}{\beta_0}\frac{1}{t_{\rm qcd}-t},
\end{align}
where the scale $\Lambda_{\rm QCD}\equiv \Lambda_0 e^{-t_{\rm qcd}}$ 
gives the scale of the infrared quantities such as the chiral order 
parameters.
Here, in order to go beyond the scale lower than $\Lambda_{\rm QCD}$, 
we introduce the infrared cutoff effect \cite{Higashijima:1983gx} 
that the gauge coupling constant stop increasing at a proper infrared scale, 
which is naturally expected by the confinement.
Adopting the cutoff scheme in Ref.~\cite{Aoki:1990eq}, the running gauge 
coupling constant is given as follows:
 \begin{align}
  \alpha_s(t)=
  \begin{cases}
   \displaystyle \frac{2\pi}{\beta_0}\frac{1}{t_{\rm qcd}-t}, & t<t_{\rm ir} 
  \\[8pt]
   \displaystyle \frac{2\pi}{\beta_0}\frac{1}{t_{\rm qcd}-t_{\rm ir}}
   +\frac{\pi}{\beta_0}\frac{(t-t_1)^2-(t_{\rm ir}-t_1)^2}{(t_{\rm ir}-t_1)(t_{\rm qcd}-t_{\rm ir})^2},
   & t_{\rm ir}<t<t_1
  \\[8pt]
   \displaystyle  \frac{2\pi}{\beta_0}\frac{1}{t_{\rm qcd}-t_{\rm ir}} 
   -\frac{\pi}{\beta_0}\frac{t_{\rm ir}-t_1}{(t_{\rm qcd}-t_{\rm ir})^2},& t_1 < t
  \end{cases},
\end{align}
where we set a fixed dimensionless scale for $t_1$ to be $t_{\rm qcd}+1$, and
$t_{\rm ir}$ is left as an infrared cutoff scale parameter.
Obeying Ref.~\cite{Aoki:1990eq}, $t_{\rm ir}$ should be parametrized by 
$\Delta_{\rm ir}$ as $t_{\rm ir}= t_{\rm qcd}-0.5\cdot (\Delta_{\rm ir}+1)$,
and we take the following parameter:
\begin{align}
 \Lambda_{\rm QCD}&=484\ {\rm MeV},\ \Delta_{\rm ir}=-0.5.
\end{align}
By the analysis using the ladder Schwinger-Dyson equation,
it is confirmed that the physical quantities  such as the chiral condensates are not sensitive to the choice of the infrared cutoff scale parameter $t_{\rm ir}$.

Finally, we discuss the initial condition of the fermion potential 
$V(\sigma;t)$.
The initial cutoff scale $\Lambda_0$ has to be large enough
so that the obtained RG flow well approximates the renormalized 
trajectory starting from the ultraviolet limit,
$\Lambda_0\rightarrow \infty$.
At the ultraviolet limit, the fermion potential vanishes because
the effective average action agrees with the bare QCD action, 
or if it exists there it would be strongly suppressed soon by its higher dimensionality.
In practical calculation,  
we take the vanishing fermion potential as the initial condition, and
set the initial cutoff scale $\Lambda_0$ to be large enough so that
the infrared quantities does not depend on the $\Lambda_0$ within a given 
numerical precision.
Actually we set $\Lambda_0$ to be the Z boson mass scale, $M_Z=91.2$ GeV.

\end{document}